\documentstyle[seceq,epsf,wrapft]{ptptex}

\newcommand{\tr}{\mbox{tr}}

\def\fsl#1{\setbox0=\hbox{$#1$}           % set a box for #1 
   \dimen0=\wd0                                 % and get its size
   \setbox1=\hbox{/} \dimen1=\wd1               % get size of /
   \ifdim\dimen0>\dimen1                        % #1 is bigger
      \rlap{\hbox to \dimen0{\hfil/\hfil}}      % so center / in box
      #1                                        % and print #1
   \else                                        % / is bigger
      \rlap{\hbox to \dimen1{\hfil$#1$\hfil}}   % so center #1
      /                                         % and print /
   \fi}                                         %

\newcommand{\arctanh}{\mbox{arctanh}}

\title{Reconsidering Composite Higgs Loop Effects \\
in the Top Mode Standard Model }

\author{Michio {\sc Hashimoto}
\footnote{E-mail: michioh@eken.phys.nagoya-u.ac.jp}
 }

\inst{
Department of Physics Nagoya University, Nagoya 464-8602 Japan}

\recdate{April 27, 1998}

\abst{
Composite Higgs loop effects in the top mode standard model are 
discussed by using the Miransky-Tanabashi-Yamawaki (MTY) approach 
based on the Schwinger-Dyson equation. 
The top mass is obtained as 179 GeV for the Planck scale cutoff 
($ \Lambda \simeq 10^{19} $ GeV). 
This result is different from that of 
the Bardeen-Hill-Lindner (BHL) approach 
based on the renormalization group equation (RGE), 
with QCD plus Higgs loop effects included
($ m_t \simeq 205 $ GeV). 
Detailed comparison of the MTY approach with the BHL approach 
is made. We derive ``RGE'' from the Pagels-Stokar formula 
by considering the infrared mass as the ``renormalization point''. 
Then, it is found that the MTY approach including 
the composite Higgs loop effects is only partially 
equivalent to the BHL approach with QCD plus Higgs loop effects. 
The difference essentially results from the treatment of the composite 
Higgs propagator, or more precisely, of $Z_H^{-1}$. 
Our results can be summarized as 
$ m_t \mbox{ (Ours)} \simeq 1/\sqrt{2} m_t \mbox{ (MTY)} $, 
in contrast to 
$ m_t \mbox{ (BHL)} \simeq \sqrt{2/3} m_t \mbox{ (MTY)} $, 
where $ m_t \mbox{ (MTY)} \simeq \mbox{250 GeV}$ is the original 
MTY prediction without Higgs loop effects. 
}

\begin{document}

\maketitle

\baselineskip 0.7cm

\setcounter{equation}{0}
\section{Introduction}

Recently, the top quark was discovered by the CDF and D0 group. 
Its mass was found to very large, approximately 174 GeV.~\cite{pdg} 
Why is the top quark so much heavier than other quarks and leptons? 
The explication of this mass hierarchy is one of the most urgent 
and interesting problems in particle physics. 
Since only the top quark mass is 
near the electro-weak symmetry breaking scale 250 GeV, 
it seems natural to think that the top quark may have an intimate 
relation to electro-weak symmetry breaking; that is 
the top quark may be connected with 
the Higgs sector in the standard model (SM). 
An idea related to this thought is that concerning 
the top quark condensate, which was proposed 
by Miransky, Tanabashi and Yamawaki (MTY)~\cite{MTY} and 
by Nambu,~\cite{nambu} before experiments revealed 
the top quark mass to be as large as it is. 
In this idea, the standard Higgs scalar is replaced by 
the corresponding bound state of the top and anti-top quarks. 
Thus the model may be called the ``top mode standard model'' (TMSM), 
in contrast to the ordinary SM using the elementary Higgs particle, 
the ``Higgs mode'' standard model. 
While the original MTY approach to the TMSM was based on 
the Schwinger-Dyson (SD) equation and 
the Pagels-Stokar (PS) formula,~\cite{PS} 
the TMSM has been further formulated elegantly 
through a renormalization group (RG) approach 
by Bardeen, Hill and Lindner (BHL)~\cite{BHL} 
using the 1-loop RG flow of the SM, 
in which the Higgs particle becomes composite 
at a scale $ \Lambda $.~\cite{eguchi} 
It is known~\cite{6} that the BHL approach including 
only QCD effects is equivalent to the MTY approach 
at $1/N_c$-leading order. 

The advantage of the TMSM is to obtain the relation of 
the electro-weak symmetry breaking scale to the top quark mass 
and the Higgs particle ($ \bar{t}t $) mass 
without introducing unknown  particles. 
In this model, however, there has been the difficulty that 
the top quark mass is predicted to be over 200 GeV. 
If we consider the ``top mode GUT''~\cite{GUT} etc., of course, 
we can bring down the top quark mass. 

However, we wish to consider whether or not the TMSM 
of the original simplest version is dead by including 
loop effects of the composite Higgs boson and the weak gauge boson. 
In the BHL approach, which is based on the perturbative RGE, 
it does not seem that the situation is changed, 
for instance, by using 2-loop RGE~\cite{2-loop} or 3-loop RGE. 
Thus we will take the original MTY approach. 
In the MTY approach, the mass function behavior at higher momentum is 
important. This means that the behavior of 
the effective top-Yukawa coupling near cutoff is described clearly. 
This is in contrast to the BHL approach in which the top-Yukawa 
in the higher momentum region is ambiguous because of a large top-Yukawa. 

In this paper, we consider the SD equation 
including composite Higgs boson loop effects in addition to 
the MTY analysis. Since the composite Higgs propagator, which was obtained 
by Appelquist, Terning and Wijewardhana,~\cite{ATW} includes 
the ladder graph of gluon, 
the behavior of the propagator 
is quite different from the usual one; i.e., the composite Higgs 
propagator acquires an extra momentum dependence of $Z_H^{-1}(p^2)$ 
[see Eq.~(\ref{cb}): 
$Z_H^{-1}(p^2) \propto 
(\ln p^2/\Lambda_{{\rm QCD}}^2)^{-1/7} - 
(\ln \Lambda^2/\Lambda_{{\rm QCD}}^2)^{-1/7}$]. 
In addition to this extra factor, the Yukawa-type vertex 
$\Gamma_s(p^2)$ also includes ladder effects [see Eq.~(\ref{3-11}) ]. 
Due to the extra factor and the Yukawa-type vertex, 
we numerically determine the top mass to be 179 GeV 
for the Planck scale cutoff ($\Lambda \simeq 10^{19}$ GeV). 
Moreover, we give the ``RGE'' for the top-Yukawa 
by using the PS formula and the SD equation, 
and clarify the relation between the MTY approach and the BHL approach. 
We should mention that to combine our ``RGE'' with 
the BHLs RGE in the small top-Yukawa region is {\it meaningless}, 
because the two methods are different. 
Our ``RGE'' flow including QCD plus Higgs loop effects damps 
more rapidly than that in the BHLs RGE. 
Thus, our top-Yukawa at the quasi-IR fixed point is 
brought down. 
{\it The difference essentially results from the treatment of $Z_H^{-1}$}. 
In our ``RGE'', the dependence of $Z_H^{-1}(p^2,M^2)$ on the 
physical momentum $p$ is different from that on the infrared 
mass $M$, which is regarded as the ``renormalization point'', 
while there is no such a distinction for $Z_H^{\overline{MS}}(\mu^2)$ 
in the $\overline{\mbox{MS}}$ scheme. 
As a result, the answer obtained in our approach is different from 
that in the BHL approach. 
Actually, if we start with the gauged Yukawa model 
by applying the improved ladder calculation to the top-Yukawa vertex, 
we find that our ``RGE'' is equivalent to that of BHLs, 
as long as we use the solution of the 1-loop RGE as the running top-Yukawa. 

This paper is organized as follows. 
In $\S$ 2, we briefly review the 
analysis of the ladder SD equation including only QCD effects, 
following to MTY.~\cite{MTY,KSY} 
Next, we consider the SD equation including 
the Higgs boson loop effects. 
Then, we introduce the non-local gauge~\cite{NLG} 
so as to be consistent with the bare vertex approximation to the SD equation. 
In $\S$ 3, we numerically analyze the SD equation for the mass function. 
In $\S$ 4 we consider the relation between the MTY approach and 
the BHL approach. 
Section 5 is devoted to summary and discussions. 

\setcounter{equation}{0}
\section{Non-local gauge}

In this section, we consider the SD equation with one-gluon-exchange 
graph plus Higgs-boson-loop effects included. 
We introduce a non-local gauge~\cite{NLG} so as to be 
consistent with the bare vertex approximation to the SD equation. 
In this gauge, the SD equation is reduced to a single equation for the 
mass function. 

Before consideration of $ SU(2)_L \times U(1)_Y $ 
flavor symmetry corresponding to the SM, 
we first consider $ U(1)_L \times U(1)_R $ flavor symmetry 
for simplicity in the $ SU(N_c) $-gauged Nambu-Jona-Lasinio (GNJL) model:
\begin{eqnarray}
{\cal L} & = & \bar{\psi} ( i \fsl{\partial} - g \fsl{A} ) \psi 
+ \frac{G}{2N_c} \left[ ( \bar{\psi} \psi )^2 + 
( \bar{\psi} i \gamma_5 \psi )^2 \right]
- \frac{1}{2} \tr ( F_{\mu\nu} F^{\mu\nu} ) , \\
& \to & \bar{\psi} ( i \fsl{\partial} - g \fsl{A} ) \psi 
-\bar{\psi} ( \sigma + i \gamma_5 \pi ) \psi 
- \frac{N_c}{2 G} ( \sigma^2 + \pi^2 ) 
- \frac{1}{2} \tr ( F_{\mu\nu} F^{\mu\nu} ) , \label{lag}
\end{eqnarray}
where we have used the auxiliary field method, 
$ \sigma = \bar{\psi}\psi $ and $ \pi = \bar{\psi} i \gamma_5 \psi $. 
Here, $ \psi $ belongs to the fundamental representation of $ SU(N_c) $, 
and $ g $ and $ G $ are the gauge coupling and the 4-fermi coupling, 
respectively. 

The simplest version of the GNJL model, the $ U(1) $-gauged NJL model 
with $ U(1)_L \times U(1)_R $ chiral symmetry, was first studied 
by Bardeen, Leung and Love in the ladder SD equation.~\cite{BLL} 
A full set of spontaneous chiral symmetry breaking solutions 
of the ladder SD equation and the critical line were discovered by 
Kondo, Mino and Yamawaki, and independently by 
Appelquist, Soldate, Takeuchi and Wijewardhana.~\cite{KMY} 
This dynamics was applied to the phenomenology, i.e., the TMSM, 
by Miransky, Tanabashi and Yamawaki.~\cite{MTY} 

We now give a brief review of the MTY result. 
We consider the SD equation for the fermion propagator 
$iS_f^{-1}(p) \equiv A(-p^2)\fsl{p}-B(-p^2)$ with a one-gluon-exchange graph, 
which is obtained from the Cornwall-Jackiw-Tomboulis (CJT) 
potential~\cite{CJT} of $ O(N_c) $ under a 2-loop approximation. 
We use the bare vertex approximation to the coupling of 
the fermion and gauge boson. If we take the Landau gauge, 
the wave function $ A(p_E^2) $ is equal to unity. 
Therefore, the Landau gauge is the most preferable in this approximation 
for consistency with the Ward-Takahashi (WT) identity.~\cite{DSBbook} 
After angular integration in Euclidean momentum, 
the SD equation for the mass function takes the form
\begin{equation}
B(x) = \sigma + 3 g^2 C_2 \int \frac{d y}{(4 \pi )^2} \; y 
\frac{B(y)}{y + B(y)^2}\frac{1}{\mbox{max($ x $,$ y $)}} , \label{ladderN}
\end{equation}
where $ C_2 = (N_c^2 -1)/2N_c $ is the quadratic Casimir constant 
of the fundamental representation, 
the chiral condensation $\sigma$ is by definition given by 
$\sigma =  \frac{g}{\Lambda^2} \int d y \; y \frac{B(y)}{y+B(y)^2}$ with 
$ g = \Lambda^2 G /4 \pi^2 $, and $x \equiv p_E^2$. 
Hereafter we use only Euclidean momentum and omit the subscript of $E$. 
Equation~(\ref{ladderN}) can be rewritten as the following differential 
equation and boundary conditions (BC's): 
\begin{eqnarray}
B''(x)  -  \frac{ ( \frac{\lambda (x)}{x})''}
{( \frac{\lambda (x)}{x} )'} B'(x)- \left( \frac{\lambda(x)}{x} \right)' 
\frac{x B(x)}{x+B^2} & = &  0 , \label{difeq1} \\
B(\Lambda^2) + \frac{1}{1+\frac{1}{\ln \Lambda^2/\Lambda_{{\rm QCD}}^2}}
\Lambda^2 B'(\Lambda^2) & = & \sigma , 
\hspace{1.8cm} \mbox{(UV-BC)} \\
x^2 B'(x) & \to & 0 \; \; (x \to 0) ,  \quad \mbox{(IR-BC)}  
\end{eqnarray}
where we have used a standard technique called the ``improved ladder'' 
calculation to take account of running effects of the gauge coupling 
in the non-Abelian gauge theory:~\cite{higashijima} 
\begin{eqnarray}
\lambda & \equiv & \frac{3 C_2 \alpha_s }{ 4\pi }
\Rightarrow \lambda(x) \theta(x-y) + \lambda(y) \theta(y-x) , \\
\lambda(x) & \equiv & \frac{c_m}{\ln (x/ \Lambda_{{\rm QCD}}^2)} , \\
c_m & \equiv & \frac{9 C_2}{11 N_c - 2 N_f} = 4/7 \; \mbox{ (for SM)} \; .
\end{eqnarray}
From (\ref{difeq1}), the mass function is obtained approximately 
as~\cite{KSY} 
\begin{equation}
B(x) \simeq M \left( \frac{ \ln x/\Lambda_{{\rm QCD}}^2}
{ \ln M^2/\Lambda_{{\rm QCD}}^2} \right)^{-c_m} , \label{2-12}
\end{equation}
where $ M $ is the infrared mass defined by $ M = B(M^2) $. 
The PS formulae with isospin breaking, which were obtained 
by MTY,~\cite{MTY} are  
\begin{eqnarray}
F_{\pi^0}^2 & = & \frac{N_c}{8\pi^2} \int_{0}^{\Lambda^2} d x x
\frac{B(x)^2 - \frac{x}{2}B(x)B'(x)}{(x+B(x)^2)^2} , \label{MTYPS1} \\
F_{\pi^{\pm}}^2 & = & \frac{N_c}{8\pi^2} \int_{0}^{\Lambda^2} d x 
\frac{B(x)^2 - \frac{x}{2} B(x)B'(x) + 
\frac{B(x)^3 B'(x)}{x+B(x)^2}}{x+B(x)^2} , \label{MTYPS2}
\end{eqnarray}
where we have assumed maximal isospin breaking ($ m_b = 0$). 
Even in this case, $ \delta \rho  = F_{\pi^{\pm}}^2/F_{\pi^0}^2 - 1 $ 
is about 2\%. 
From $ F_{\pi} = 246 $ GeV, MTY predicted the top mass as 250 GeV 
with cutoff $\Lambda = 10^{19} $ GeV. 
%%
%% Figure 1
%%
\begin{figure}[t]
  \begin{center}
    \leavevmode
     \epsfxsize= 14cm \epsfbox{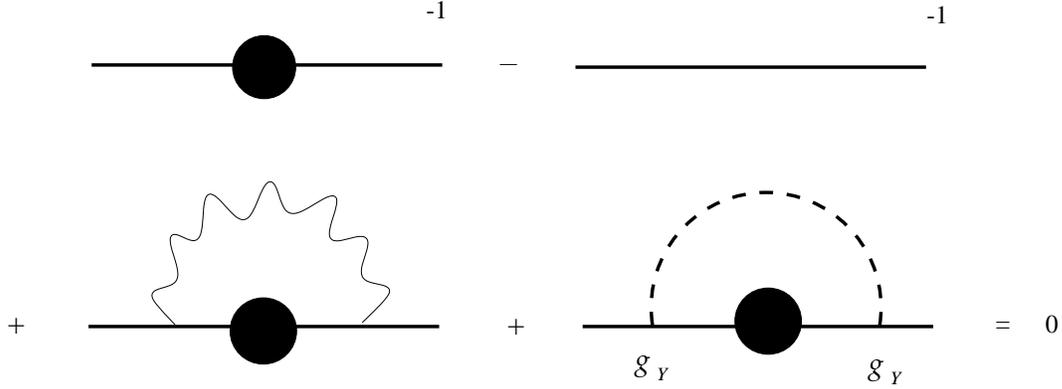}%
  \end{center}
\caption{The Schwinger-Dyson equation. 
The solid line with the shaded blob, the solid line without the shaded blob, 
the wavy line, and the dotted line 
represent the full fermion propagator $ S_f $, 
the bare fermion propagator, 
the bare gauge boson 
propagator $ D_{\mu\nu} $, and the composite Higgs propagator $ D_{H}$, 
respectively. Note that the bare fermion propagator inverse 
to the momentum $ p $ is equivalent 
to $ \fsl{p}  - \sigma $ in the auxiliary field method. }
\label{fig1}
\end{figure}

Recently, the top quark was discovered and 
the mass was determined to be about 174 GeV, 
which is somewhat smaller than the MTY value, though on the order of 
the weak scale, as predicted by MTY. 
Thus, we consider the SD equation with one-gluon-exchange 
graph plus Higgs-loop effects (Fig.~1). 
The SD equation for Fig.~1 is given as follows:
\begin{eqnarray}
A (p^2) = 1 &+& \frac{g^2 C_2}{p^2}\int \frac{d k^4}{(2\pi)^4} 
\frac{A(k^2)}{A(k^2)^2 k^2 + B(k^2)^2} 
\left[ \frac{p \cdot k}{q^2} + 2\frac{(p \cdot q)(k \cdot q)}{(q^2)^2}
\right] \nonumber \\
& + &\frac{g^2 C_2}{p^2}\int \frac{d k^4}{(2\pi)^4} 
\frac{A(k^2)\xi^{-1}}{A(k^2)^2 k^2 + B(k^2)^2} \left[ 
\frac{p \cdot k}{q^2} - 2\frac{(p \cdot q)(k \cdot q)}{(q^2)^2}
\right] \nonumber \\
& - & \frac{1}{p^2} 
\int \frac{d k^4}{(2\pi)^4} \left[ 
\frac{A(k^2) p \cdot k}{A(k^2)^2 k^2 + B(k^2)^2}
\sum_{\sigma, \pi} D_{H}(q^2) \right] , \label{A} \\
B (p^2) = \sigma &+&  g^2 C_2\int \frac{d k^4}{(2\pi)^4} 
\left[ (3+\xi^{-1})\frac{B(k^2)}{A(k^2)^2  k^2 + B(k^2)^2}
\right] \nonumber \\
& + & \int \frac{d k^4}{(2\pi)^4} 
\left[ \frac{B(k^2)}{A(k^2)^2 k^2 + B(k^2)^2}
\left( D_{\sigma}(q^2)-D_{\pi}(q^2)\right) \right] ,  \label{B}  
\end{eqnarray}
where we have made the bare vertex approximation, 
$ g_Y  = 1 , \; i \gamma_5  \mbox{ for $\sigma$, $\pi$}$, 
and $ D_{\mu\nu}(p)$ and 
$ D_{H}(p) $ are the bare gauge boson propagator
$(D_{\mu \nu}(q)  =  \frac{1}{q^2} ( g_{\mu \nu} - 
(1-\xi^{-1}) \frac{q_\mu q_\nu}{q^2} ))$ and 
the composite Higgs propagator $(H=\sigma,\pi)$, respectively. 

Since we take the bare vertex approximation, 
we need to set $ A(p^2)=1 $ for consistency with the WT 
identity. Of course, the coupled SD equations of the wave function 
and the mass function could be considered under a suitable vertex ansatz. 
We introduce, instead of consideration of such SD equations, 
the non-local gauge $\xi^{-1}(q^2)$ so as to fix $ A(p^2)=1 $ 
consistently with the bare vertex approximation. 
From this standpoint, Eqs.~(\ref{A}) and (\ref{B}) are reduced 
to a single equation for the mass function $ B(p^2) $ 
by requiring $A(p^2)=1$ through the freedom 
of gauge choice.~\cite{NLG} 
In this gauge, $ B(p^2) $ becomes the very mass function. 

It is  well known that the Landau gauge $(\xi^{-1}=0)$ gives 
$ A(p^2)=1 $ in the analysis of the one-gluon-exchange graph; 
i.e., the second term and the third term in the r.h.s. of 
Eq.~(\ref{A}) are canceled out. 
We consider the following trick to reparametrize 
the integrating momentum:
\begin{eqnarray}
\hspace{-1cm} \lefteqn{ 0=\int \frac{d k^4 }{(2\pi)^4}
\frac{1}{k^2 + B(k^2)^2} \left[ \frac{p \cdot k}{q^2}
+ 2 \frac{(p \cdot q)( k \cdot q)}{(q^2)^2}  \right], }  \\
& \simeq & 
 \int \frac{d k'^4 }{(2\pi)^4}   
\frac{1}{q'^2 + B(q'^2)^2} \left[ \frac{p \cdot q'}{k'^2}
+ 2 \frac{(p \cdot k')(q' \cdot k')}{(k'^2)^2} \right] , \label{NLG1}
\end{eqnarray}
where we have assumed the momentum-shift-invariant regularization and 
$q(q') \equiv p-k(k')$. 
By using the relation of Eq.~(\ref{NLG1}), 
we can rewrite Eq.~(\ref{A}) as 
\begin{eqnarray}
\lefteqn{p^2 (A(p^2)-1)} \nonumber \\
= & & g^2 C_2 \int \frac{d k^4}{(2\pi)^4} 
\frac{1}{k^2 + B(k^2)^2} \left[ -2 (p \cdot q)(k \cdot q) 
\frac{\xi^{-1}(q^2)}{(q^2)^2} + p \cdot k \frac{\xi^{-1}(q^2)}{q^2} \right] 
\nonumber \\
& - & \sum_{\sigma, \pi } \int 
\frac{d k^4}{(2\pi)^4} \frac{p \cdot k}{k^2 + B(k^2)^2} D_{H}(q^2) , 
\label{2-28} \\
\simeq & & 
g^2 C_2  \int \frac{d k'^4}{(2\pi)^4} 
\frac{1}{q'^2 + B(q'^2)^2} \left[ -2 (p \cdot k')(q' \cdot k') 
\frac{\xi^{-1}(k'^2)}{(k'^2)^2} + p \cdot q'
\frac{\xi^{-1}(k'^2)}{k'^2} \right] \nonumber \\
& - & \sum_{\sigma, \pi} 
\int \frac{d k'^4}{(2\pi)^4} \frac{p \cdot q'}{q'^2 + B(q'^2)^2} 
D_{H}(k'^2) , \label{2-29} \\
= & & 
\int \frac{y' d y' d \Omega_k'}{(4\pi)^2} 
\frac{ p \cdot q'}{q'^2 + B(q'^2)^2} 
\left[
2 g^2 C_2 \frac{\xi^{-1}(y')}{y'} 
- \sum_{\sigma , \pi} 
D_{H}(y') \right] ,  \label{2-25}
\end{eqnarray}
where $ y' \equiv k'^2$ and 
we have set $ A(p^2)=1 $ already on the r.h.s. of Eq.~(\ref{2-28}). 
In addition, Eq.~(\ref{2-29}) has been obtained by shifting 
the integrating momentum from $k$ to $q'$. 
We find the non-local gauge $ \xi (y) $ 
by setting r.h.s. to zero:
\begin{equation}
2g^2 C_2 \frac{\xi^{-1}(y)}{y} = \sum_{\sigma, \pi} D_{H}(y) .
\label{NLG2} 
\end{equation}

When we derived Eq.~(\ref{NLG1}), we used a momentum-shift-invariant 
regularization, for instance, the dimensional regularization. 
Of course, the naive cutoff regularization is not invariant under 
shifting the integrating momentum. 
If we consider the constant mass function $B(x) = m $ and 
a finite cutoff $\Lambda$, 
the r.h.s. of Eq.~(\ref{NLG1}) is obviously not equal to zero. 
Thus one might suspect whether our non-local gauge $\xi(x)$ is 
consistent with $A(x) \simeq 1$ for finite cutoff. 
By substituting Eq.~(\ref{NLG2}) into Eq.~(\ref{2-28}), we obtain 
the wave function $A(x)$ as follows:
\begin{equation}
x( A(x)-1)=\int \frac{dk^4}{(2 \pi)^4} \frac{1}{k^2 + B(k^2)^2} 
\left[ - \frac{(p \cdot q)( k \cdot q)}{q^2} -\frac{p \cdot k}{2} \right]
\sum_{\sigma, \pi} D_{H}(q^2) . \label{consistencyA}
\end{equation}
In the case that we take the composite Higgs propagator 
$D_{H}(q^2)$ as the linear-$\sigma$ model type or the NJL type
(see (\ref{3-31})), 
we can confirm $A(x) \simeq 1$, assuming that the scalar mass is very small 
compared with the cutoff $\Lambda$. 
Finally, we may consider the non-local gauge of Eq.~(\ref{NLG2}) to be 
consistent with $A(x) \simeq 1 $. 
If we take the naive cutoff regularization from the beginning, 
such a problem would not occur. 
In Ref.~\citen{unpublished}, the SD equations~(\ref{A}) and (\ref{B}) are 
considered in the non-local gauge in such a case 
for the gauged Yukawa model. 
However, the analysis there is very complicated. 

We substitute the non-local gauge of Eq.~(\ref{NLG2}) into Eq.~(\ref{B}). 
Then, we obtain the integral equation for the mass function as
\begin{eqnarray}
B(x) 
& \simeq & \sigma + 3 g^2 C_2 \int \frac{d k^4}{(2\pi)^4} 
\frac{B(k^2)}{k^2 + B^2}\frac{1}{q^2} + 
\frac{1}{2} \int \frac{d k'^4}{(2\pi)^4} 
\frac{B(q'^2)}{q'^2 + B(q'^2)^2} \sum_{\sigma, \pi} D_{H}(k'^2)
 \nonumber \\
& & \quad + \int \frac{d k'^4}{(2\pi)^4} \frac{B(q'^2)}{q'^2 + B(q'^2)^2} 
( D_{\sigma}(k'^2) - D_{\pi}(k'^2)) , \label{2-34} \\
& \simeq & \sigma + \frac{3 \alpha_{\mbox{s}} C_2}{4 \pi} 
\int d y \; y \; \frac{B(y)}{y + B(y)^2}\frac{1}{\mbox{max}(x,y)} 
\nonumber \\ 
& & \quad + \frac{1}{32 \pi^2} \int d y' \; y' \; \frac{B(y)}{y' + B(y)^2} 
\frac{y'}{\mbox{max}(x,y')} \sum_{\sigma, \pi} D_{H}(y') \nonumber \\ 
& & \quad + \frac{1}{16 \pi^2} 
\int d y' \; y' \; \frac{B(y')}{y' + B(y')^2} \frac{y'}{\mbox{max}(x,y')}
( \; D_{\sigma}(y') - D_{\pi}(y') \; ) , \label{B2}
\end{eqnarray}
where Eq.~(\ref{2-34}) has been obtained by using the non-local gauge 
of Eq.~(\ref{NLG2}) after shifting the integrating momentum 
from $k$ to $q'$. Note that to derive Eq.~(\ref{B2}) 
we have used the following trick for the angular integral: 
\begin{equation}
\int \frac{d k^4}{(2\pi)^4} 
\frac{B(k^2)}{k^2 + B(k^2)^2}\frac{1}{q^2} \simeq   
\int \frac{d k'^4}{(2\pi)^4} 
\frac{B(q'^2)}{q'^2 + B(q'^2)^2}\frac{1}{k'^2} ,
\end{equation}
\begin{equation}
\mbox{i. e., } \hspace{0.5cm} 
\int d \Omega'_k \frac{B(q'^2)}{q'^2 + B(q'^2)^2} 
\simeq 
\frac{y' B(y')}{y' + B(y')^2}\frac{1}{\mbox{max}(x,y')} . 
\end{equation}

\setcounter{equation}{0}
\section{Numerical analysis of the Schwinger-Dyson equation \\
including Higgs-Loop effects}
 
In this section, we analyze Eq.~(\ref{B2}) using two approaches, 
one in which the composite Higgs propagator is taken as the NJL type
[Case I], 
and one in which we use the composite Higgs propagator obtained by 
ladder $ 1/N_c $-leading analysis~\cite{ATW} 
(i.e., the NJL-type vertex at $1/N_c$-leading order plus 
the gauge-boson-ladder graph included)[Case II]. 

Since we are interested in the high momentum behavior of $ B(x) $, 
we may neglect the Higgs mass; i.e., 
we assume $ D_{\sigma}(y') \simeq D_{\pi^0}(y') $. 

First, we take the composite Higgs propagator 
$ D_{H}(p^2) $ as the NJL model propagator [Case I] 
for comparison of our analysis with the BHL approach including 
only QCD plus Higgs loop effects. The NJL-type propagator 
in the high momentum region is given approximately by 
\begin{eqnarray}
\lefteqn{\hspace*{-1.5cm} 
D_{H}^{-1}(x) - D_{H}^{-1}(0) } \nonumber \\
&  & \hspace{-0.7cm} = -
 \frac{N_c}{8\pi^2} x
 \left[
 \ln\left( 1 + \frac{\Lambda^2}{\sigma^2} \right) - I(x, \sigma^2)
 + \frac{2\Lambda^2 + 4\sigma^2 +x}{4\Lambda^2 + 4\sigma^2 +x}
 I(x, \sigma^2+\Lambda^2) \right] , \\
& & \hspace{-0.7cm} \simeq - 
\frac{N_c}{8 \pi^2} x ( - \ln \frac{x}{\Lambda^2} + r ) , \label{3-31}\\[3mm]
& & \hspace*{-1.5cm} I(x,z) \equiv 2 \sqrt{\frac{x+4 z}{x}} 
\arctanh{\sqrt{\frac{x}{x+4 z}}}  , \\
& &\hspace*{-0.6cm} r \equiv 
 \frac{6}{\sqrt{5}} \arctanh \sqrt{\frac{1}{5}} \simeq  1.29123 , 
\end{eqnarray}
where we have neglected fermion condensation $ \sigma $ 
$(x \gg \sigma)$. 

For $ SU(2)_L \times U(1)_Y $ flavor symmetry, 
we simply replace $ \sum_{\sigma, \pi }$ in Eq.~(\ref{B2}) by 
$ \sum_{\sigma, \pi^0, \pi^+ }$. 
Of course, the fermion propagator takes the form 
$ iS_f^{-1}(p) = A(-p^2) \fsl{p} + A_5 (-p^2) \gamma_5 \fsl{p} - B(-p^2) $ 
under consideration of $ SU(2)_L $ symmetry. 
The pseudoscalar mass function $ B_5(-p^2) $ can always be 
rotated away by the chiral symmetry, but $ A_5(-p^2) $ cannot. 
We discuss this problem later. 
In any case, we now continue the analysis for Eq.~(\ref{B2}). 

By using the improved ladder calculation, 
the bifurcation method~\cite{bifurcation}, and the PS formula, 
we obtain the fermion mass as 221 GeV for the cutoff 
$ \Lambda = 10^{19}$ GeV. 
We will not describe this result in more detail, 
because this analysis is made in a parallel manner 
to the following analysis. 
This result is stable with respect to changes in $ r $. 
If we vary $ r = 0 \sim 2 $, the mass is 219 $ \sim $ 222 GeV. 

On the other hand, in the BHL approach 
with QCD plus Higgs loop effects (without $SU(2)_L \times U(1)_Y$ 
gauge loop effects), the top-Yukawa is obtained as
\begin{equation}
Y (t) \equiv \frac{1}{y_t^2} = \frac{N_c + \frac{3}{2}}{16 \pi^2}
\frac{1}{2 c_m -1} \left( t( \mu^2 ) - t^{2c_m}( \mu^2 ) 
t^{1-2c_m}( \Lambda^2 ) \right) , \label{BHLFpi}
\end{equation}
where $ t( \mu^2 ) \equiv \ln \mu^2 / \Lambda_{{\rm QCD}}^2 $. 
This top-Yukawa yields a top mass of 205 GeV. 
Thus, it seems that the MTY approach including the loop effects of 
the NJL-type propagator is {\it not} equivalent to the BHL approach. 
In next section, 
we will discuss the relation in detail. 
%%
%% Figure 2
%%
\begin{figure}[b]
  \begin{center}
    \leavevmode
     \epsfxsize=14cm \epsfbox{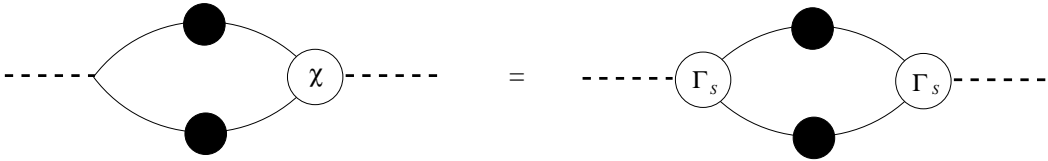}%
  \end{center}
\caption{The composite Higgs propagator inverse including the Yukawa-type 
vertex $\Gamma_s$. The solid line with the shaded blob, 
the dotted line, $ \chi $, and $ \Gamma_s $ represent the full fermion 
propagator, the composite Higgs, Bethe-Salpeter amplitude, and 
the Yukawa-type vertex at zero momentum transfer, respectively. }
\label{fig2}
\end{figure}

Next, we consider [Case II], using 
the composite Higgs propagator obtained 
by Appelquist, Terning and Wijewardhana,~\cite{ATW} (see Fig.~2). 
In the improved ladder calculation of $ 1/N_c $-leading analysis, 
we find 
\begin{eqnarray}
\lefteqn{ D_{H}^{-1}(x) - D_{H}^{-1}(0) } \hspace{0.7cm} \nonumber \\
& \simeq & 
\frac{N_c}{8\pi^2}\int_0^{\Lambda^2} d y \Gamma_s^2 (y) \left[ 
( \frac{y}{x}-2 \; ) \theta (x-y) - \frac{x}{y} \theta (y-x) \right] ,\\
& = & - \frac{N_c }{8\pi^2} x 
\frac{ t(\Lambda^2)^{2c_m} }{2c_m-1} \left[ 
\int_0^{t(x) -t(M^2)} \hspace{-0.5cm} 2 (t-u)^{1-2c_m} e^{-u} (1-e^{-u}) \; du
\; \, - t(\Lambda^2)^{1-2c_m} \right. \nonumber \\
& & \hspace{1.0cm}\left.  
- t(M^2)^{1-2c_m}( \frac{2M^2}{x}-\frac{M^4}{x^2}) 
\right] + 
\frac{N_c}{8\pi^2}\int_0^{M^2} d y \Gamma_s^2 (y) 
( \; \frac{y}{x}-2 \; ) , \label{D1} \\ 
& \simeq & 
- \frac{N_c}{8 \pi^2} x \frac{t(\Lambda^2)^{2c_m}}{2c_m -1} 
\left( t(x)^{1-2c_m} - t(\Lambda^2)^{1-2c_m} \right) ,
\hspace{0.5cm} (x \gg M^2) \label{cb} \\ 
t(x) & = & \ln x/\Lambda_{{\rm QCD}}^2 , \\
\Gamma_s (x) & \equiv & \frac{d B(x)}{d \sigma} , \quad 
\mbox{(The Yukawa-type vertex at zero momentum transfer)}  \label{3-11} \\
& \simeq & \left( \frac{t(x)}{t(\Lambda^2)} \right)^{-c_m}, \quad ( x > M^2 )
\end{eqnarray}
where $ M $ is the infrared mass to normalize the mass function, 
and we have neglected the third, fourth and fifth terms of Eq.~(\ref{D1}), 
because these terms are $ O(M^2) \ll x $. 
%%%%%%%%%%%%%%%%%%%%
%% Figure 3
%%%%%%%%%%%%%%%%%%%%
\begin{figure}[b]
  \begin{center}
    \leavevmode
     \epsfxsize=14cm \epsfbox{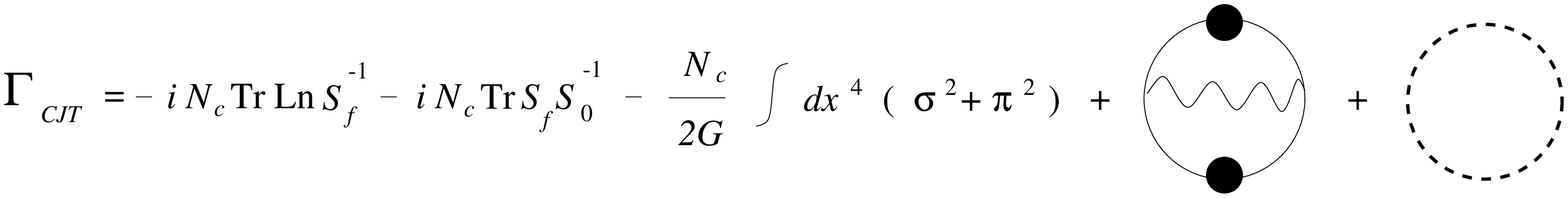}%
  \end{center}
\caption{The CJT potential. The solid line with the shaded blob, 
the dotted line, and the wavy line represent 
the full fermion propagator $ S_f$, 
the composite Higgs propagator, and the bare 
gauge boson propagator, respectively. 
In the second term, $ S_0^{-1}= \fsl{p} - \sigma $ is the bare fermion 
propagator. 
The last term in this potential 
is $ O(N_c^0) $, and the other terms are $ O(N_c) $. 
Note that the composite Higgs propagator is given by Fig.~2. }
\label{fig3}
\end{figure}
%%%%%%%%%%%%%%%
%%
%%%%%%%%%%%%%%

Moreover, in the present case we need to modify Eq.~(\ref{B2}). 
If we start with the CJT potential of Fig.~3, 
we find that the two $ g_Y $'s of Fig.~1 must 
be replaced by $ \Gamma_s $ (see Fig.~4). 
We can confirm this easily by differentiating the CJT potential 
with respect to the full fermion propagator $S_f$, noting that 
the composite Higgs propagator in Case~II consists of 
the ladder graph (Fig.~2). 
Because we consider the composite Higgs propagator inverse as 
the r.h.s. of Fig.~2, which includes two Yukawa-type vertices of 
$\Gamma_s$ in our approximation, 
the SD equation does not take the usual form with 
one bare vertex and one 1PI full-vertex. Instead, it takes 
the form of Fig.~4 with two Yukawa-type vertices of $\Gamma_s$. 
%%%%%%%%%%%%%%%%%%%%
%% Figure 4
%%%%%%%%%%%%%%%%%%%%
\begin{figure}[b]
  \begin{center}
    \leavevmode
     \epsfxsize=14cm \epsfbox{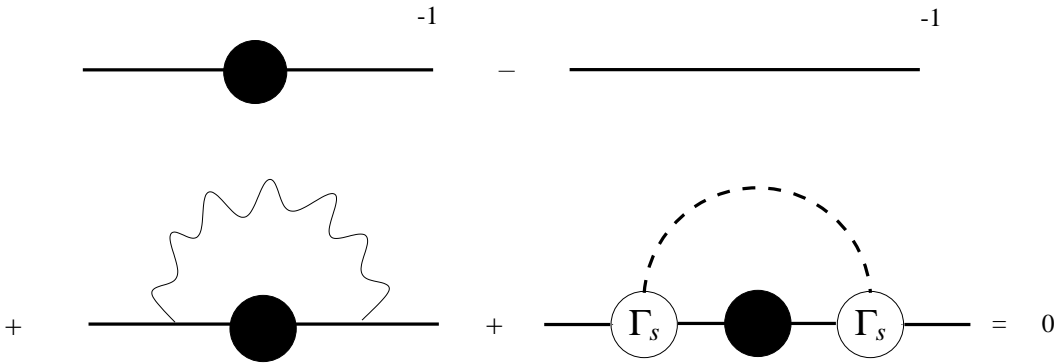}%
  \end{center}
\caption{The Schwinger-Dyson equation including the composite Higgs loop 
effects. The solid line with the shaded blob, 
the solid line without the shaded blob, the dotted line, 
and the wavy line represent the full fermion propagator, 
the bare fermion propagator, the composite Higgs propagator, 
and the bare gauge boson propagator, respectively. 
Note that two $\Gamma_s$ vertices are used instead of one bare vertex 
and one 1PI-full vertex, because 
the composite Higgs propagator is given approximately 
by the r.h.s. of Fig.~2. }
\label{fig4}
\end{figure}
Finally, we obtain the SD equation for the mass function as 
\begin{eqnarray}
B(x) & \simeq & \sigma + \int d y \; y \; 
\frac{B(y)}{y + B(y)^2} \left[ \frac{\lambda(x)}{x} \theta (x-y) + 
\frac{\lambda(y)}{y} \theta (y-x) \right] \nonumber \\ 
& & + \frac{1}{32 \pi^2} \int d y \; y \; \frac{B(y)}{y + B(y)^2} 
\frac{y \Gamma_s (x) \Gamma_s (y)}{\mbox{max}(x,y)} 
\sum_{\sigma, \pi^0, \pi^+ } D_{H}(y) . \label{B3}
\end{eqnarray}
In this expression, the divergence of $ t( \Lambda^2 )^{2c_{m}} $ 
from $ D_{H}^{-1}(y) $ is canceled out by the same one 
from the two $ \Gamma_s $, and the result does not depend on whether 
we use the expression of $\Gamma_s (x)^2$ or $\Gamma_s (y)^2 $ 
in place of $ \Gamma_s (x) \Gamma_s (y)$ in Eq.~(\ref{B3}). 
The differences between the top mass predictions are about 1 GeV 
in these cases. 

We can solve Eq.~(\ref{B3}) simply by using the bifurcation method. 
Then we can show that the linearized 
differential equation corresponding to Eq.~(\ref{B3}) is not second order 
but third order: 
\begin{eqnarray}
\lefteqn{ \hspace{-0.8cm}
\frac{d^3}{d t^3}B(t)  +  \frac{\Delta_2}{\Delta_1}\frac{d^2}{d t^2}B(t)
+ \left[ \frac{\Delta_3}{\Delta_1} + c_m (\frac{1}{t} + \frac{1}{t^2})
- \frac{3}{4 N_c}(1+\frac{c_m}{t}) \frac{t^{-2c_m}}{K(t)} \right]
\frac{d}{d t} B(t) } \hspace{1.5cm} \nonumber \\[2mm]
& & \hspace{-2.3cm} + 
\left[ \frac{\Delta_4}{\Delta_1}\frac{c_m}{t} 
- \frac{3}{4 N_c} \left( \frac{\Delta_5}{\Delta_1} 
- \frac{c_m}{t} - \frac{c_m^2}{t^2} \right) \frac{t^{-2c_m}}{K(t)}
+ \frac{3}{4 N_c}(1+\frac{c_m}{t}) 
\frac{K'(t)}{K(t)}\frac{t^{-2c_m}}{K(t)} \right] B(t) \nonumber \\[3mm]
& & \hspace{9.0cm} = 0 , \label{3-15} 
\end{eqnarray}
\begin{eqnarray}
&&\hspace{-0.8cm}
\Delta_1 \equiv 1 + \frac{c_m + 2}{t} + \frac{c_m}{t^2}, \\
&&\hspace{-0.8cm}
\Delta_2 \equiv 2 + \frac{3c_m + 6}{t} +
          \frac{c_m^2+7c_m+6}{t^2} + \frac{c_m^2+4c_m}{t^3}, \\
&&\hspace{-0.8cm}
\Delta_3 \equiv 1 + \frac{4c_m+4}{t} + \frac{2c_m^2-13c_m+6}{t^2}
          + \frac{2c_m^2+8c_m+18}{t^3}+\frac{2c_m^2+2c_m}{t^4}, \\
&&\hspace{-0.8cm}
\Delta_4 \equiv 1 + \frac{2c_m+3}{t} + \frac{c_m^2+6c_m+2}{t^2}
          + \frac{2c_m^2+4c_m-2}{t^3} + \frac{c_m^2}{t^4}, \\
&&\hspace{-0.8cm}
\Delta_5 \equiv 1 - \frac{4c_m-7}{t} - \frac{c_m^2+4c_m-6}{t^2}
          - \frac{c_m^3 + 2c_m^2 -2c_m}{t^3}-\frac{c_m^3-2c_m^2}{t^4}, \\
&&\hspace{-1.1cm}
K(t) \equiv \frac{1}{2c_m-1}\left( 
          t^{1-2c_m}(x) - t^{1-2c_m}(\Lambda^2) \right), \label{3-21} \\
&&\hspace{-1.4cm}
B(t) \; \; \to \, M , \quad 
         \mbox{($ t \to t(M^2) = \ln M^2/\Lambda_{{\rm QCD}}^2 $ )}, \\
&&\hspace{-1.4cm}
B'(t) \; \to \; \, 0, \quad \;
          \mbox{($ t \to t(M^2) = \ln M^2/\Lambda_{{\rm QCD}}^2 $ )}, \\
&&\hspace{-1.4cm}
B''(t) \: \to \; 0 , \quad \;
           \mbox{($ t \to t(M^2) = \ln M^2/\Lambda_{{\rm QCD}}^2 $ )}, \\
&&\hspace{-1.4cm}
B'''(t) \to \; 0 , \quad \;
            \mbox{($ t \to t(M^2) = \ln M^2/\Lambda_{{\rm QCD}}^2 $ )}. 
\end{eqnarray}
By using the analytical expression of the PS formula,~\cite{marciano} 
which neglects $ B'(x) $ and replaces 
the denominator $ x + B(x)^2 $ by $ x $ in Eqs.~(\ref{MTYPS1}) and 
(\ref{MTYPS2}), 
\begin{equation}
F_{\pi}^2 = \frac{N_c}{8\pi^2} \int_{M^2}^{\Lambda^2} d x 
\frac{B(x)^2}{x} ,
\end{equation}
we numerically obtained the top quark mass $ m_t = 179 $ GeV with 
$ \Lambda = 10^{19}$ GeV and $ F_{\pi} = 246 $ GeV. 
We obtain the Table I for various cutoffs. 

The differential equation Eq.~(\ref{3-15}) is complicated, however; 
the main term comes from $\Gamma_s (x)^2 Z_H(x)$ in Eq.~(\ref{B3}), 
where we define $D_H^{-1}(x) \equiv - Z_H^{-1}(x) (x+M_H^2(x))$, with 
$Z_H^{-1}(x) $ being given from Eq.~(\ref{cb}): 
\begin{equation}
 Z_H^{-1}(x) \simeq 
\frac{N_c}{8 \pi^2}\frac{(\ln \Lambda^2/\Lambda_{{\rm QCD}}^2)^{2c_m}}{2c_m-1}
           \left( ( \ln x/\Lambda_{{\rm QCD}}^2)^{1-2c_m}-
                  (\ln \Lambda^2/\Lambda_{{\rm QCD}}^2)^{1-2c_m} \right) \; .
\end{equation}
This factor of $\Gamma_s(x)^2 Z_H(x)$ blows up more rapidly 
than the one in Case~I. 
Thus, the mass function in Case II grows more in the high momentum region 
than that in Case~I, and as a result the prediction 
for the top mass is smaller. 
In the next section, the relation between the  MTY approach and 
the BHL approach is described in detail. 

\vspace{1cm}

\renewcommand{\arraystretch}{1.5}
%%
%% Table 1
%%
\begin{table}[h]
\begin{center}
\begin{tabular*}{130mm}[c]{c|@{\extracolsep{\fill}}cccccccc} \hline 
 $ \Lambda $ & $ 10^{21} $ & $ 10^{20}$ & $ 10^{19}$  & 
$ 10^{18} $  & $ 10^{17} $ & $ 10^{16}$ & $ 10^{15} $ & 
$ 10^{14} $ \\ \hline 
$ m_t $ & 175  & 177  & 179  & 181  & 184  & 
187  & 190 & 194  \\ \hline
\end{tabular*}
\end{center}
\vspace{0.5cm}
\caption{The top mass for various cutoffs (GeV) in Case II. }
\label{tab1}
\end{table}

\setcounter{equation}{0}
\section{The Relation between the Miransky-Tanabashi-Yamawaki approach 
and the Bardeen-Hill-Lindner approach}

Now we consider the relation between the MTY approach and the BHL approach. 
In the previous section, we found numerically that our approach is 
{\it not} precisely equivalent to the BHL approach 
in two cases for $D_H(p^2)$. 
Thus, we wish to obtain an analytical relation. 
From the bifurcation method and the analytical PS formula, 
we find generally 
\begin{equation}
F_{\pi}^2 (M^2) = \frac{N_c}{8\pi^2} \int_{M^2}^{\Lambda^2} d x M^2 
\frac{\frac{f(x)^2}{f(M^2)^2}}{x} , \label{PS2}
\end{equation} 
where $ f(x) $ is a dominant solution to the SD equation for the mass function 
and $ B(x) = M f(x)/f(M^2) $. 
Needless to say, the mass function cannot be divided into a one variable 
function like $f(x)$ under consideration of sub-dominant solutions. 
In fact, the mass function becomes $B(x)=M f(x,M^2)$ in this case, 
where~$f(M^2,M^2)=1$. 
In the analysis of the one-gluon-exchange graph, 
for instance, $ f(x) $ is nearly equal to 
$ ( \ln x/\Lambda_{{\rm QCD}}^2 )^{-c_m} $ from Eq.~(\ref{2-12}). 
If we read $ M $ as a ``renormalization point'' $ \tilde{\mu} $ 
in Eq.~(\ref{PS2}), we can define a ``Yukawa coupling'' corresponding 
to the BHL approach as~\cite{6}
\begin{equation}
Y(\tilde{\mu}) = 1/y_t^2 \equiv 
                 \frac{F_{\pi}^2(\tilde{\mu}^2)}{2\tilde{\mu}^2} = 
 \frac{N_c}{16 \pi^2} \int_{\tilde{\mu}^2}^{\Lambda^2} d x  
\frac{\frac{f(x)^2}{f(\tilde{\mu}^2)^2}}{x} . \label{yukawa}
\end{equation}
From Eq.~(\ref{yukawa}), we obtain an ``RGE'' for 
the ``Yukawa coupling'' in the MTY approach as follows:
\begin{equation}
\hspace{-3.5cm} \mbox{[``RGE{\tiny MTY}'']} \qquad
\frac{d Y}{d \tilde{t}} = - \frac{N_c}{16 \pi^2} - 
                            \frac{2 f'(\tilde{t})}{f(\tilde{t})}Y , 
                          \qquad  \tilde{t} \equiv \ln \tilde{\mu}^2 \; . 
\label{MTYyukawa}
\end{equation}
We should mention that $Y$ becomes justly zero when 
$\tilde{\mu} \to \Lambda$ in Eq.~(\ref{yukawa}). 
This corresponds to the compositeness condition of the BHL approach. 
On the other hand, we know the 1-loop RGE of the SM 
for Yukawa coupling, 
\begin{eqnarray}
\lefteqn{\hspace{-1.2cm} \mbox{[ RGE{\tiny BHL} ]}} \nonumber \\ 
&&\frac{d Y}{d t} = - \frac{N_c + 3/2}{16 \pi^2} + 
\frac{1}{4 \pi } \left( 3\frac{N_c^2-1}{N_c} \alpha_s + 9/4 \alpha_2 
+ 17/12 \alpha_1 \right) Y , 
\label{BHLyukawa}
\end{eqnarray}
where $t = t(\mu^2) = \ln \mu^2$, $\mu$ is the renormalization point in 
the $\overline{\mbox{MS}}$ scheme, 
the Higgs loop effects give the factor $3/2$, and 
$ \alpha_{1} $ and $ \alpha_{2} $ are $ U(1)_Y$- and $SU(2)_L$-gauge 
couplings, respectively. 
In the case of $U(1)_L \times U(1)_R $ flavor symmetry, 
i.e., Eq.~(\ref{lag}), 
the 1-loop RGE is given by
\begin{equation}
\hspace{-2cm} \mbox{[ RGE$_{U(1)_L \times U(1)_R}$ ]} \qquad 
\frac{d Y}{d t} = - \frac{N_c + 1}{16 \pi^2} + 
\frac{1}{4 \pi } \left( 3\frac{N_c^2-1}{N_c} \alpha_s \right) Y . 
\label{ULUR}
\end{equation}
``RGE{\tiny MTY}'' is similar to RGE{\tiny BHL} or 
RGE$_{U(1)_L \times U(1)_R}$ , and in the fact 
they become identical in the large $N_c$ limit.~\cite{6} 
However, their meanings are different. 
In RGE{\tiny BHL}, because of using a perturbative RGE, 
the flow of large $ y_t $ in the high energy region is ambiguous. 
On the other hand, the mass function $ f(t) $ in ``RGE{\tiny MTY}'' 
is given clearly at higher momentum rather than at low energy. 
In other words, ``RGE{\tiny MTY}'' is more reliable than 
RGE{\tiny BHL} in the large $ y_t $ region. 
We may understand ``RGE{\tiny MTY}'' as a ``non-perturbative RGE'' 
in a sense. 
We should not mix the two approaches; for instance, 
we should not combine ``RGE{\tiny MTY}'' with RGE{\tiny BHL} in 
the small top-Yukawa region, because these approaches are based on 
different manners of thinking. 

The differential equation for $f(x)$ obtained from 
Eq.~(\ref{B3}) is given approximately as follows:
\begin{equation}
\frac{d^2 f}{d t^2} + \frac{d f}{d t} + \left( \lambda (t) 
- \frac{3}{4} \frac{2c_m -1}{N_c} 
\frac{1}{t - t^{2c_m}t(\Lambda^2)^{1-2c_m}} \right) f = 0 . \label{B4} 
\end{equation}
For Eq.~(\ref{B4}), it can be shown numerically 
that $ f'' $ is almost irrelevant. 
We may regard $t(x)$ as $\tilde{t}=t(\tilde{\mu}^2)$, because $f(x)$ is 
a one-variable function. 
Thus, ``RGE{\tiny MTY}'' in Case~II becomes 
\begin{eqnarray}
\lefteqn{\hspace{-1.5cm} \mbox{[``RGE{\tiny MTY}'' in Case II ]}} \nonumber \\
& & \frac{d Y}{d \tilde{t}} = - \frac{N_c}{16 \pi^2} + 
\left( 2 \lambda (\tilde{t}) 
- \frac{3}{2} \frac{2c_m -1}{N_c} 
\frac{1}{\tilde{t} - \tilde{t}^{2c_m}t(\Lambda^2)^{1-2c_m}} \right) Y . 
                                                   \label{MTYyukawaC}
\end{eqnarray}
In the same way, by use of the NJL-type propagator (\ref{3-31}), 
``RGE{\tiny MTY}'' in Case I can be found as
\begin{eqnarray}
\mbox{\hspace{-2.2cm} [``RGE{\tiny MTY}'' in Case I ]} \nonumber \\
& & \hspace{-2.3cm} \frac{d Y}{d \tilde{t}} = - \frac{N_c}{16 \pi^2} + 
\left( 2 \lambda (\tilde{t}) 
- \frac{3}{2N_c}\frac{1}{-\tilde{t} + 
         \ln \Lambda^2/\Lambda_{{\rm QCD}}^2+r} \right) Y . 
                                                    \label{MTYyukawaNJL}
\end{eqnarray}
To be general, we obtain ``RGE{\tiny MTY}'' with QCD plus the composite 
Higgs loop effects [Case III] from Eq.~(\ref{B3}) by using 
$\Gamma_s(x)^2 Z_H(x)$ as follows:
\begin{eqnarray}
\mbox{\hspace{-3cm} 
[``RGE{\tiny MTY}'' in Case III (the general case)]} 
                                                        \nonumber \\
& & \hspace{-5.5cm} 
\frac{d Y}{d \tilde{t}} = - \frac{N_c}{16 \pi^2} + 
2 \left( \lambda (\tilde{t}) 
- \frac{1}{32 \pi^2} \sum_H \Gamma_s(\tilde{t})^2 Z_H(\tilde{t}) \right) Y ,
\label{MTYyukawa_gen}
\end{eqnarray}
where we have assumed that $f''$ is negligible. 
Note that Eq.~(\ref{MTYyukawa_gen}) takes the same form even if 
we consider $U(1)_L \times U(1)_R$ flavor symmetry, 
which is discussed in $\S$~2: The term of $\Gamma_s(x)^2 Z_H(x)$ is 
the same as the above one for $H=\sigma, \pi^0$. 

If we start with the gauged Yukawa model using 
the improved ladder calculation to take account of 
the running effects of the top-Yukawa, we come to substitute 
$Z_H(\tilde{t}) \Gamma_s(\tilde{t})^2 = y_t^{{\rm sol}}(\tilde{t})^2 /2$ 
into Eq.~(\ref{MTYyukawa_gen}), 
where $y_t^{{\rm sol}}$ is the solution of the 1-loop RGE. 
Then, we find that Eq.~(\ref{MTYyukawa_gen}) in this case is 
just equal to RGE{\tiny BHL} up to $SU(2)_L \times U(1)_Y$ 
gauge contributions. 

In our model, however, $y_t(x)^2$ is not necessarily equal to 
$\tilde{y}_t(x)^2 \equiv 2 Z_H(x) \Gamma_s(x)^2$, where the factor of 
$2$ arises from our normalization. 
By definition, $y_t(x)$ is the top-Yukawa at zero momentum transfer, 
while $\tilde{y}_t(x)$ can be interpreted as the top-Higgs coupling 
at the same momentum of the Higgs boson as that of the top quark. 
Thus, these physical meanings are different. 
We should mention 
\begin{equation}
 F_{\pi^{a}}^2 = Z_{\pi^{a}}^{-1}(0,M^2) \sigma^2 , \qquad (a=0,\pm)
                                                         \label{4-9}
\end{equation}
which is derived from the WT identity for the axial-vector vertex 
including the auxiliary field.~\cite{WThiggs} 
In Eq.~(\ref{4-9}), we have written explicitly the infrared mass dependence 
of $Z_{\pi^{a}}$. Hereafter, we do not distinguish $\pi^{a}$, 
because we have neglected the deviation of $F_{\pi^0}$ 
from $F_{\pi^{\pm}}$ in this paper. 
By the definition (\ref{yukawa}), another expression for our $y_t$ 
is obtained as follows:
\begin{eqnarray}
 y_t(M^2)^2 & \equiv & Z_{\pi}(0,M^2)\frac{2M^2}{\sigma^2} , \label{4-10} \\
       & \simeq & 2 Z_{\pi}(0,M^2) \frac{f(M^2)^2}{f(\Lambda^2)^2}, 
                                                         \label{4-11}        
\end{eqnarray}
where Eq.~(\ref{4-11}) was derived from UV-BC of 
$\sigma \simeq B(\Lambda^2)$ for the mass function. 
On the other hand, we find 
\begin{eqnarray}
 \tilde{y}_t(x)^2 & \equiv & 2\Gamma_s(x,M^2)^2 Z_H(x,M^2) , \label{4-12} \\
                  & \simeq & 2 Z_{\pi}(x,0) \frac{f(x)^2}{f(\Lambda^2)^2},
                                                          \label{4-13}
\end{eqnarray}
where we have neglected the $M^2$ dependence of $Z_{H}$ 
in the high-energy region, and we have used 
$Z_{H}(x,0) = Z_{\pi}(x,0) \; \; [H=\sigma,\pi]$ due to 
the chiral symmetry. 
Thus, the deviation of $\tilde{y}_t$ from $y_t$ results essentially 
from that of $Z_{\pi}(M^2,0)$ from $Z_{\pi}(0,M^2)$. 
Of course, we cannot estimate $Z_{\pi}(0,M^2)$, 
as long as the bifurcation method is used. 
Generally speaking, it is very difficult to obtain the behavior of 
the mass function around zero momentum under consideration of 
the running coupling effects in the SD approach. 
%%%%%%%%%%%%%%%%%%%%
%% Figure 5
%%%%%%%%%%%%%%%%%%%%
\begin{wrapfigure}{r}{6.6cm}
     \epsfxsize=7.5cm 
\epsfbox{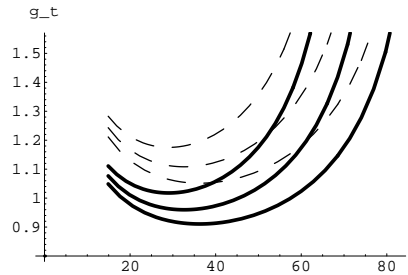}%
\caption{The RGE flow. The dotted line and the solid line 
represent the BHL approach and our approach in Case II, respectively. 
From top to bottom, $\Lambda = 10^{15}, 10^{17} \mbox{ and } 10^{19}$GeV. } 
\label{fig5}
\end{wrapfigure}

In contrast to this, there is no such a distinction for 
$Z_{H}^{\overline{MS}}(\mu^2)$ in the $\overline{\mbox{MS}}$ scheme. 
This is the reason why $\tilde{y}_t(\mu^2)$ becomes equivalent to 
$y_t(\mu^2)$ in the $\overline{\mbox{MS}}$ scheme. 
We should mention that {\it this point is not an artifact of the 
$1/N_c $-expansion}. 
The same conclusion can be also be drawn for the case of 
$U(1)_L \times U(1)_R $ flavor symmetry, where $A_5 (x)$ disappears. 
Thus, we find that the difference between our result and that of 
the BHL is not due to the ambiguity of $A_5 (x)$. 

Actually, our result of Eq.~(\ref{MTYyukawaC}) 
%and (\ref{MTYyukawaNJL}) are 
is different from that of BHL. 
Due to this difference, the RGE flow is changed (see Fig.~5). 

Finally, 
we obtain analytical expressions of the decay constant 
in our approach as follows:
\begin{eqnarray}
&& \hspace{-1.3cm} F_{\pi}^2 (M^2) = \frac{N_c}{8\pi^2} M^2 t(M^2)^{2c_m}
                         (-\ln M^2/\Lambda^2 + r)^{1/2} \times 
                                                        \nonumber \\
& &  \hspace{0.8cm} \int_{t(M^2)}^{t(\Lambda^2)}
      t^{-2c_m}(-t+\ln \Lambda^2/\Lambda_{{\rm QCD}}^2 + r)^{-1/2} , 
\quad \mbox{[ Ours in Case I ]}                            \label{analFpiNJL}
\end{eqnarray}
\vspace{0.1cm}
\begin{equation}
F_{\pi}^2 (M^2) = \frac{\eta N_c}{8\pi^2}\frac{M^2}{2c_m-1}
                      \left[ t(M^2) - t(M^2)^{2c_m} t(\Lambda^2)^{1-2c_m} 
                      \right] ,   \hspace{-0.5cm} \label{analFpi} 
\end{equation}
\begin{equation}
\hspace{-0.0cm} \mbox{where } 
\quad \eta  =  \left\{
\begin{array}{cl}
1 , & \quad \mbox{[ MTY/BHL including only QCD effects ]} \\
\frac{3}{2} , & \quad \mbox{[ BHL including QCD plus Higgs loops effects ]} \\
2 , &  \quad \mbox{[ Ours in Case II ]} 
\end{array}
\right. \label{4-17}
\end{equation}
for $ N_c = 3 $. 
The analytical expressions of $ F_{\pi} $ in Eqs.~(\ref{analFpi}) 
and (\ref{4-17}) give $ m_t$ (Ours) $\simeq$ $1/\sqrt{2} m_t $ 
(MTY) and $ m_t $ (BHL) $\simeq$ $\sqrt{2/3} m_t $ (MTY), 
where $m_t$ (MTY) $\simeq$ 250 GeV stands for the original 
MTY value, corresponding to $\eta=1$ in Eq.~(\ref{4-17}).  

Now, we comment on the technique of $1/N_c$-sub-leading analysis, 
because the $1/N_c$-expansion is partially used in our approach. 

We have to notice that the mass function $ B(x) $ must not be written 
naively as a series in $ 1/N_c $.~\cite{NTL} Actually, 
one might be tempted to expand the mass function $ B(x) $ 
as $ B(x) = B_0 (x) + \frac{1}{N_c} B_1 (x)+ \cdot \cdot \cdot $. 
Then, one would have 
$ B_0 (x) = - \frac{1}{N_c} B_1 (x)$ on a critical line 
in the second order phase transition. For consistency, 
we would find $B_0(x)=B_1(x)=0$ on the critical line. 
Then, the critical coupling would not be changed 
by including any higher order $1/N_c$-corrections. 
Of course, we disagree with this claim.~\footnote{
However, such an expansion 
of the chiral condensation $\sigma$ is considered in
Ref.~\citen{cvetic}.} 
This means that $ 1/N_c $-expansion 
of the order parameter should not be done naively. 
In Ref.~\citen{NTL}, this point will be discussed in detail.

\setcounter{equation}{0}
\section{Summary and discussions}

We have found that the top quark mass can be brought down 
below 200 GeV (in our analysis (Case II) 179 GeV for Planck scale cutoff), 
by using the SD equation with QCD plus 
the composite Higgs loop effects. 
This can be understood analytically as $ m_t \mbox{ (Ours)} \simeq 
\frac{1}{\sqrt{2}} m_t \mbox{ (MTY)}$. 
It was suggested that the difference of the result between our approach 
and the BHL approach reflects the different treatment of $Z_H^{-1}$. 
We should note that the top mass is increased by 
about 10\% when $ SU(2)_L \times U(1)_Y $-gauge loop effects are switched. 

However, it must be mentioned that the composite Higgs propagator 
(\ref{cb}) may be ambiguous at higher momentum 
because the technique in Ref.~\citen{ATW} is based on resummation of 
the Taylor series around zero momentum of the Higgs boson. 
Recently, the composite Higgs boson propagator was obtained analytically 
under some assumptions without use of 
the resummation technique.~\cite{Gusynin} 
The expression for this propagator was obtained only in the case of 
constant gauge coupling. In that work, the coefficient of 
$(x/\Lambda^2)^{\frac{-1+\sqrt{1-4\lambda}}{2}}$ 
in the composite Higgs propagator is smaller than 
that of Ref.~\citen{ATW}. If the situation is the same in 
the case of the improved ladder analysis, 
the top mass may be brought down more. 
 
Of course, some difficulties may be pointed out technically 
to our approach at sight. 
In particular, as in the previous discussion, 
we need to take the fermion propagator 
$ i S_f^{-1} (p) = A(-p^2) \fsl{p} + A_5 (-p^2) \gamma_5 \fsl{p} -B(-p^2) $ 
with consideration of $ SU(2)_L $ symmetry. 
This problem seems to be serious. 
It is expected that $ A_5 (-p^2)$ can vanish 
if we choose a good non-local gauge for the $ SU(2)_L $ gauge. 
Then, we will be able to make a full analysis for the TMSM 
in the MTY approach. That is a future work.

\section*{Acknowledgements}

The author is very grateful to K. Yamawaki for 
helpful discussions and reading the manuscript. 
Thanks are also due to K.-I. Kondo and M. Tanabashi 
for providing an unpublished note~\cite{unpublished} 
and due to M. Lindner for discussions.

\end{document}